# Interferometric optical mass measurement in the low-reference regime

*Jan Becker* [*]

## Abstract

The precise optical, label-free, measurement of mass at the nanoscale has been significantly advanced by techniques based on interferometric scattering, such as mass photometry (MP). These methods exploit the interference between a scattered and reference field to achieve a high signal-to-noise ratio (SNR) for weakly scattering objects (e.g. proteins) and are currently limited to masses $\geq$ 40 kDa. Standard MP employs a mask that attenuates the reference field, allowing for the increase in illumination power without saturation of the detector. In this theoretical study, we examine how the SNR evolves when extending reference attenuation beyond conventional levels: entering the low-reference regime. Our simplified model finds that a substantial SNR enhancement can be achieved when the magnitude of reference matches that of the scattered field and investigate refractive index tuning as a potential method to reach the required attenuation in practice. The accomplishable SNR improvement can be tailored to a given mass region, i.e. allowing the detection of masses < 40 kDa.

*Keywords:* interferometric scattering, mass photometry, dark field, refractive index tuning, mode matching, optical nulling

## Introduction

Inferring mass of biological systems from optical, label-free, measurements has recently become a powerful addition to traditional techniques, such as mass spectrometry [1]. The current most accurate method to achieve this is based on interferometric-scattering (iScat) [2] and was developed as a successor to darkfield (DF) [3] microscopy, the standard method to detect nano-scale objects. In both cases the scatterer is illuminated with a plane wave and sits on top of a glass coverslip, within some medium. The glass-medium interface reflects parts of the illumination, which together with the scattered component can interfere on the detector (e.g. a camera; iScat). In DF this interference is prevented by adding a mask in the back-focal-plane (BFP) which fully absorbs the reference beam.

The reason why DF could not be successfully applied to the quantification of small scatterers, such as proteins, is that the achievable signal-to-noise ratio (SNR) is too low. iScat circumvents this by boosting the SNR through the interference of the scattered with a reference field (i.e. removing the mask in a DF setup). However, iScat is still limited with respect to (w.r.t.) the reachable SNR of even smaller proteins, due to the finite full-well-depth (FWD) of the detector. To further boost SNR, the mask is re-introduced, but this time instead of fully blocking, it is attenuating. Which together with a simultaneous increase of illumination power achieves the inference of protein mass $m \geq$ 40 kDa, a technique termed mass photometry (MP) [4], where proteins bind to the glass coverslip over time. However, for smaller proteins $m <$ 40 kDa the possible SNR in standard MP, again, is non-sufficient.

---

[*] *Orcid: 0000-0001-8529-4244*

We note the historical improvements in terms of SNR which were: use a ... 1) mask which fully *absorbs* the reference field (= DF); 2) mask that fully *transmits* the reference field (= iScat); 3) mask that *partially transmits* the reference field (= MP). The main questions our work asks is: *How does the SNR level change when further decreasing the reference field beyond that of standard MP, but keeping it higher than in DF mode?*

In practice such a strong attenuation of the reference field has not yet been achieved due to the mask acting as a high-pass filter, which inevitably increases the modulation of the speckle-like raw signal introduced by the roughness of the glass coverslip [4,5]. This will then fill the FWD, making the mass quantification impossible. Hence, we investigate a different way to reduce the reference field: tuning the refractive index of the medium $n_m$ to be close to that of the glass coverslip $n_g$. This not only reduces the speckle-like background but also leads to an improved SNR, as we will show later.

In the following we develop a simple model that describes the theoretical dependency of the interferometric signal on $n_m$ and show that an optimum in terms of SNR is achieved when reference and scattered field strength are matched. Which should allow to boost the performance of standard MP at any given mass region. Further we discuss potential limitations in real experimental scenarios.

<u>Theory</u>

The measurable total photon count $N_{total}$ of the previously described interferometric measurement device is given as (note that we omit any spatial dependency):

$$N_{total} = \underbrace{N_{ref}}_{\text{reference}} + \underbrace{N_{scat}}_{\text{scattering}} + \underbrace{2 \cdot \sqrt{N_{ref} \cdot N_{scat}} \cdot \cos \Delta\varphi}_{\text{interferometric}}$$

With the detailed expression for reference $N_{ref}$ and scattered $N_{ref}$ photon counts given as eq. S120 and S121 in [5] and their mutual phase difference $\Delta\varphi$.

In the following we will analyse the change of $N_{ref}$ and $N_{scat}$ when modifying $n_m$. For the reference field we get:

$$N_{ref}(n_m) \propto \tau \cdot R(n_m)$$

With $R$ being the (power) reflection coefficient and $\tau$ the (power) transmission coefficient of the mask (typically $\tau$ = 0.1%, only affects $N_{ref}$; [5]). The reflection coefficient at normal incidence is:

$$R(n_m) = \left| \frac{n_g - n_m}{n_g + n_m} \right|^2$$

Note that making $n_m = n_g$ will yield no reflection, i.e. our mechanism to arbitrarily reduce the reference field. For the scattering component we find:

$$N_{scat}(n_m) \propto T_2(n_m) \cdot T_1(n_m) \cdot \sigma(n_m)$$

With $T_{1,2}$ being the (power) transmission coefficient for the forward & backward passing through the $n_g/n_m$ interface (at normal incidence; $n_g = 1.515$). Both transmission coefficients are given as:

$$T_1(n_m) \propto \left| \frac{2 \cdot n_g}{n_g + n_m} \right|^2$$

$$T_2(n_m) \propto \left| \frac{2 \cdot n_m}{n_g + n_m} \right|^2$$

The scattering signal is proportional to the scattering cross-section $\sigma$, which for particles much smaller than the wavelength (e.g. 445 nm; [5]) is given as:

$$\sigma(n_m) \propto n_m^4 \cdot \left| \frac{n_p^2 - n_m^2}{n_p^2 + 2 \cdot n_m^2} \right|^2 \cdot m^2$$

Note the strong dependence on $n_m$: increasing $n_m$ should strongly increase the scattering up until close to the point where $n_m \sim n_p$ (being the refractive index of the particle/protein; here $n_p = 1.46$).

We will employ the typical contrast definition used in MP, given as (eq. S117 in [5]):

$$contrast(n_m) = \frac{N_{total}(n_m) - N_{ref}(n_m)}{N_{ref}(n_m)}$$

I.e. the relative difference between the photon counts with ($N_{total}$) and without scatterer ($N_{ref}$).

The variance of the extractable signal is, according to eq. S87 in [5], given as:

$$variance(n_m) = \left[ \frac{N_{total}(n_m)}{N_{ref}(n_m)} \right]^2 \cdot \left[ \frac{1}{N_{total}(n_m)} + \frac{1}{N_{ref}(n_m)} \right]$$

And the achievable signal-to-noise ratio $snr$ is defined as (eq. S119 [5]):

$$snr(n_m) = \frac{contrast(n_m)}{\sqrt{variance(n_m)}}$$

In the following we will numerically analyse the distributions of $contrast$ & $snr$ when varying $n_m$.

<u>Results</u>

We begin our analysis by plotting the photon counts for the reference, scattering and interferometric components for BSA (bovine serum albumin) with a 0.1% mask (same as in [5]), over varying refractive index tuning $\Delta n$:

$$n_m = n_{m0} + \Delta n = 1.333 + \Delta n$$

The results are depicted in Fig. 1 top and show that the individual signal levels differ by orders of magnitude (reference >> interferometric >> scattering) at $\Delta n = 0$, making the inference of mass through the interferometric term much more suitable for small particles [2,4]. We observe two poles at $\Delta n = 0.127$ ($n_m = n_p$) and $\Delta n = 0.184$ ($n_m = n_g$), where the latter indicates a region where all three components (reference, scattered & interferometric) become roughly equal in strength.

When plotting the extractable $signal = \text{N}_{\text{total}} - \text{N}_{\text{ref}}$ we observe that we can reach $signal = N_{ref} \rightarrow N_{scat} = N_{ref}$ for a phase difference $\Delta\varphi = \pi$ and a specific refractive index $n'_m = n_{m0} + \Delta n'$ (Fig. 1 bottom; red dashed). Next, we compute the contrast enhancement w.r.t. the $\Delta n = 0$ case (Fig. 2 top) and observe a very strong peak at $n_m = n_g$. This peak does not yield a strong $snr$ as it is accompanied by even stronger noise (Fig. 2 bottom). At $\Delta n'$, however, the noise reaches a minimum.

When computing the potential SNR improvement (Fig. 3 top) we observe a peak at $\Delta n'$, due to the strongly reduced noise, while maintaining a moderate contrast enhancement. The SNR improvement can reach up to 1000-fold in the idealized case we have described so far. Note that such an enhancement requires a precise refractive index tuning, as the enhancement peak is quite narrow.

These SNR improvements depend on the condition $N_{scat}(m) = N_{ref}$, i.e. do vary with the scatterer strength. Figure 3 bottom shows the potential SNR improvement for a scatterer equivalent to 10x BSA (or BSA with a 0.001% mask). A much larger potential SNR enhancement can be realized, while the necessary $\Delta n$ to reach that peak improvement also changes.

We will now relate our findings to the mass measurement of proteins and define a reference mass $m_0$, for which we set $N_{scat}(m_0) = N_{ref}$ by tuning the refractive index to $n_m'$. In this case the total photon count is given as:

$$\text{N}_{\text{total}}(m, m_0) = \text{N}_{\text{scat}}(m_0) + \left(\frac{m}{m_0}\right)^2 \cdot \text{N}_{\text{scat}}(m_0) - 2 \cdot \sqrt{\text{N}_{\text{scat}}(m_0) \cdot \left(\frac{m}{m_0}\right)^2 \cdot \text{N}_{\text{scat}}(m_0)} =$$
$$= \text{N}_{\text{scat}}(m_0) \cdot \left[1 + \left(\frac{m}{m_0}\right)^2 - 2 \cdot \frac{m}{m_0}\right] \quad \xrightarrow{m=m_0} \quad 0$$

Yielding a photon count of zero when $m = m_0$ and only when the spatial distribution of $\text{N}_{\text{ref}}$ & $\text{N}_{\text{scat}}$ perfectly cancel each other (true in our simplified model, but not in real experiments due to the required mode matching (MM) between $\text{N}_{\text{ref}}$ & $\text{N}_{\text{scat}}$ + unwanted background; see discussion).

The optical contrast is then given as:

$$contrast(m, m_0) = \left(\frac{m}{m_0}\right)^2 - 2 \cdot \frac{m}{m_0} \quad \xrightarrow{m=m_0} \quad -1$$

Yielding a quadratic dependency and a value of -1 when observing the reference structure ($m = m_0$).

The variance of the signal is given as (see appendix for the derivation):

$$variance(m, m_0) \propto [1 + contrast(m, m_0)] \cdot [2 + contrast(m, m_0)] \quad \xrightarrow{m=m_0} \quad 0$$

Indicating the ability to achieve (in principle) noise-free inference of the mass of the reference structure. The respective SNR then becomes infinite according to:

$$snr(m, m_0) \propto \frac{1}{\sqrt{\left[1 + \frac{1}{contrast(m, m_0)}\right] \cdot \left[1 + \frac{2}{contrast(m, m_0)}\right]}} \quad \xrightarrow{m=m_0} \quad \frac{-1}{\sqrt{0}} = -\infty$$

The curves for *contrast* and *snr* are shown in Fig. 4 for $m_0$ = 10 kDa. Compared to standard MP the contrast-to-mass curve becomes quadratic instead of linear (note the linearization when $m \gg m_0$). Which means that proteins smaller and bigger than $m_0$ can generate the same optical contrast, i.e. there is an ambiguity in terms of inferring mass from optical contrast.

In terms of SNR, we observe a strong improvement (blue curve) in a region around $m_0$, indicating the ability to specifically enhance the accuracy of mass inference in the low-reference regime. Note, however, that there is also a blind spot at $m = 2 \cdot m_0$, where the SNR drops sharply. In comparison to the $\Delta n = 0$ -case (magenta) we observe a reduced SNR for $m > m_0$. Which does not cause any disadvantage in practice since the resulting SNR levels for these larger masses are anyways well above the detection limit (SNR = 3; [5]) of standard MP.

Discussion

In our work we investigate the ability to infer mass from optical interference signals with the reference field being in the low-reference regime. We found that it is possible to get substantial improvements in terms of SNR when the reference matches the scattered field. This idea is not new as it was previously already described as *optical nulling* in [6] and further discussed in [7] without the context of mass measurements. In principle, this approach (inference through perfect cancellation of the scattered signal) requires full control over the spatial magnitude and phase distribution of the reference signal, i.e. would require preparing the reference as that of a point scatterer with variable strength to enable perfect MM (note: an interesting approach incorporating better MM is [8]). Which means that perfect optical nulling is not achievable through refractive index tuning alone, albeit working in the low-reference regime might still create significant SNR improvements in practice.

The refractive index tuning method requires the scattered and reference field to be out-of-phase ($\Delta \varphi = \pi$) and the latter to be strongly attenuated. This attenuation can be potentially realized in many other ways, e.g. stronger mask + perfectly flat substrate, illuminating close to Brewster angle, etc. Depending on the amount of attenuation, the SNR of different mass regions can be enhanced, while others get slightly reduced. Hence, MP in the low-reference regime lends itself towards a different way of measuring mass altogether: instead of trying to infer within a larger mass region, working in the low-reference regime automatically picks out a much narrower subset and asks a boolean-type question (*"Is the measured mass = $m_0$?"* instead of *"What is the measured mass?"*). Which might open a new way of inferring mass from optical signals: changing the ratio reference/scattered field (which effectively changes $m_0$) and observing when events become detectable with *contrast* $\sim -1$.

Our work does not include the influence of unwanted background, which is inevitable in practice and assumes that the refractive index changes of the medium to not alter the biological properties of the scatterers (when using refractive index tuning to achieve the strong reference attenuation). Both these effects reduce the applicability of our proposed scheme and will require further experimental improvements to allow for working in the low-reference regime to become successful.

Note that despite all of this, simply the reduction of the speckle-like raw signal (caused by the glass roughness) through increasing $n_m$ might already yield substantial SNR improvement, as using a stronger attenuation mask (0.1% → 0.001%) becomes possible. Without the need to employ the boolean-inference strategy which was previously described. I.e., the future of mass measurements based on optical interference might lie in the low-reference regime.

<u>Appendix</u>

The variance is given as:

$$variance(m, m_0) = \left[1 + \left(\frac{m}{m_0}\right)^2 - 2 \cdot \frac{m}{m_0}\right]^2 \cdot \frac{1}{N_{scat}(m_0)} \cdot \left[\frac{1}{1 + \left(\frac{m}{m_0}\right)^2 - 2 \cdot \frac{m}{m_0}} + 1\right]$$

$$= \frac{1}{N_{scat}(m_0)} \cdot \left[1 + \left(\frac{m}{m_0}\right)^2 - 2 \cdot \frac{m}{m_0}\right] \cdot \left[2 + \left(\frac{m}{m_0}\right)^2 - 2 \cdot \frac{m}{m_0}\right]$$

Which we rewrite into:

$$variance(m, m_0) = \frac{1}{N_{scat}(m_0)} \cdot [1 + contrast(m, m_0)] \cdot [2 + contrast(m, m_0)] \xrightarrow{m = m_0} 0$$

$$= \frac{1}{N_{scat}(m_0)} \cdot contrast(m, m_0)^2 \cdot \left[1 + \frac{1}{contrast(m, m_0)}\right] \cdot \left[1 + \frac{2}{contrast(m, m_0)}\right]$$

<u>References</u>

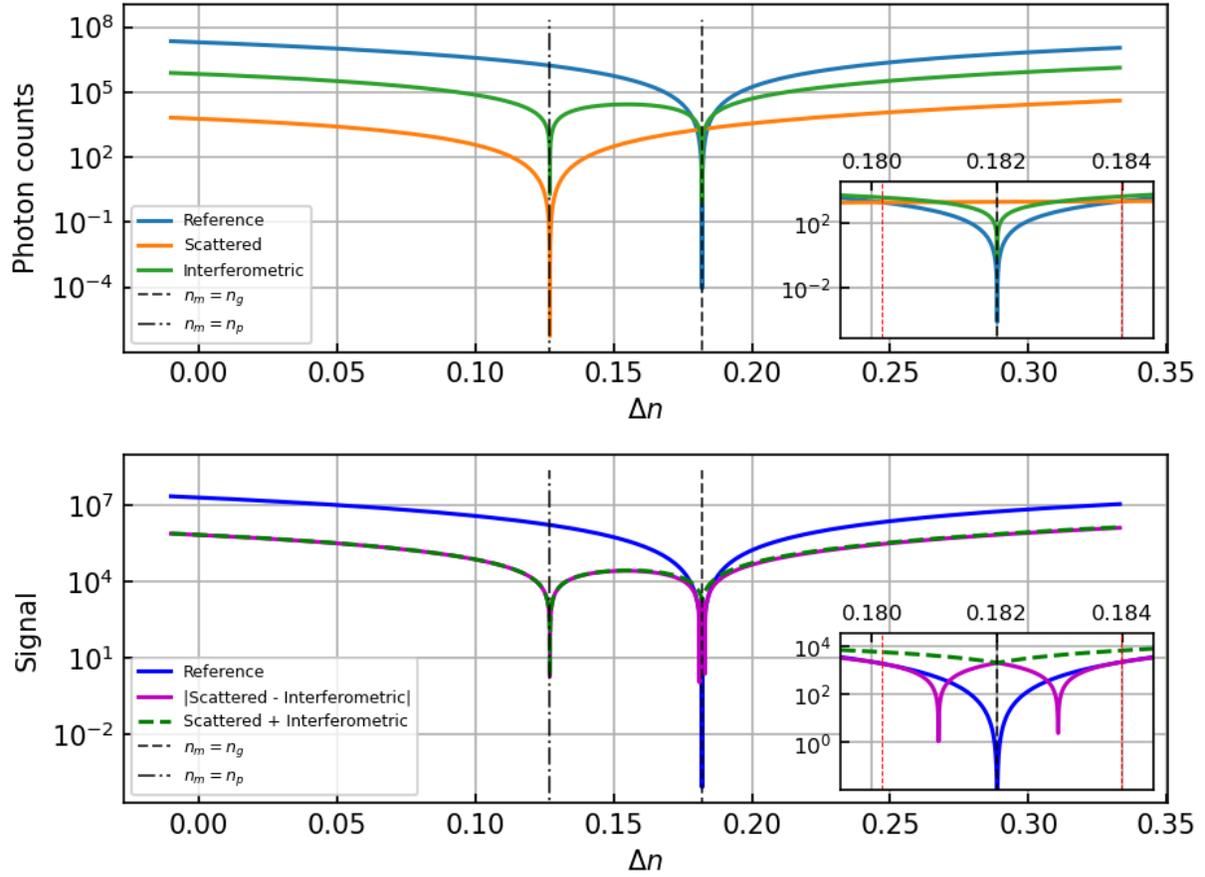

*Figure 1:* **Change of photon counts and interferometric signal through refractive index tuning.**
*Top: Simulated photon counts for the reference (blue), scattered (orange) and interferometric (green) signals for BSA ($n_p = 1.46$; $\Delta n = 0.127$) with a 0.1% mask. The inset shows the three signals around $n_g = 1.515$ ($\Delta n = 0.182$), highlighting the point where the reference and purely scattered signals are equal (red dashed). Bottom: Simulated signals when removing the reference (blue) contribution. Note that the correct sign of the interferometric signal (i.e. $\Delta \varphi = \pi$) allows to set the extractable signal (magenta) equal to the reference signal (highlighted with red dashed vertical lines).*

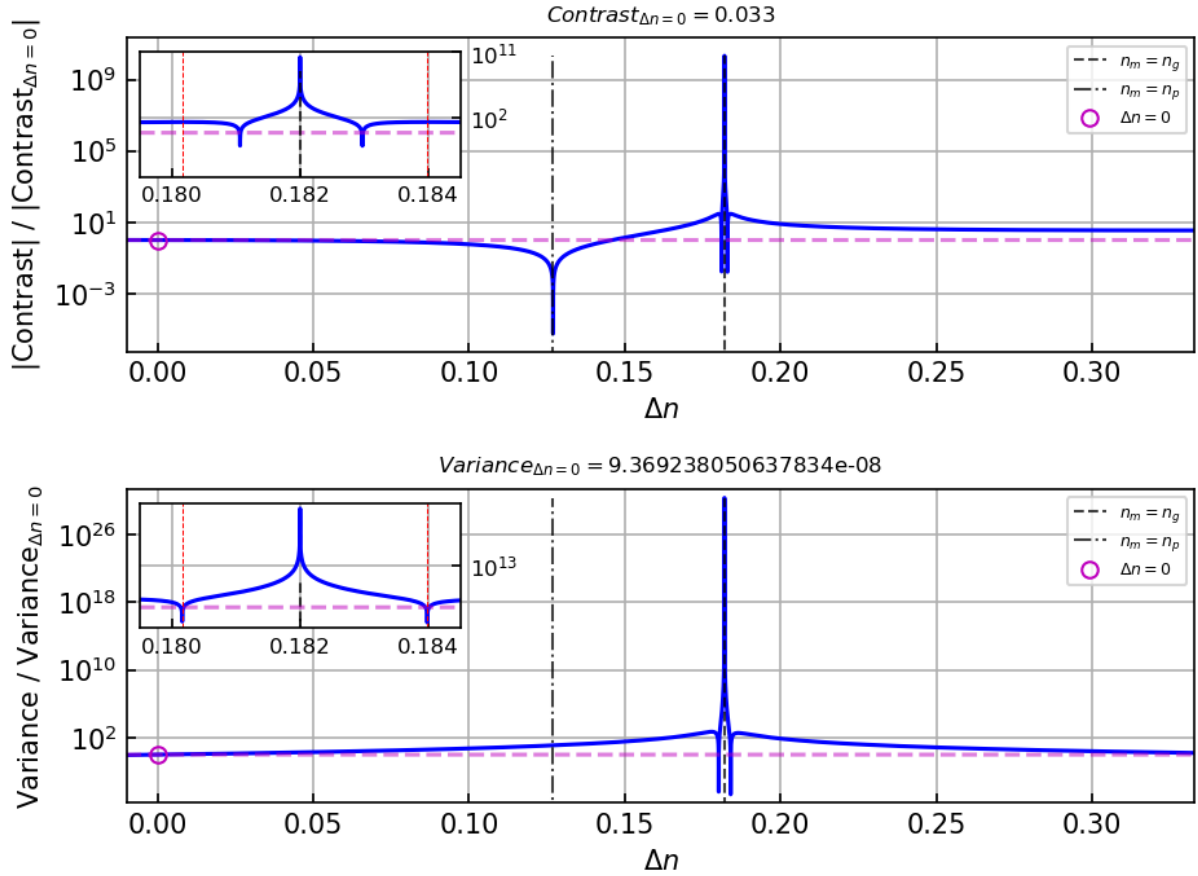

*Figure 2*: **Contrast and variance change through refractive index tuning.** *Top: Simulated contrast improvement w.r.t. no refractive index tuning ($n_{m0} = 1.333$; $\Delta n = 0$). A large contrast improvement can be generated at $n_m = n_g$, albeit at the cost of increased noise (see variance below) due to a drastically reduced reference component (see blue curve in Fig. 1). Bottom: Achievable variance improvement also w.r.t. the $\Delta n = 0$-case. Note the reduction in noise at the point where reference and scattered component are equal (red dashed), which comes with a modest increase in terms of optical contrast (see top), yielding an overall SNR improvement (see Fig. 3).*

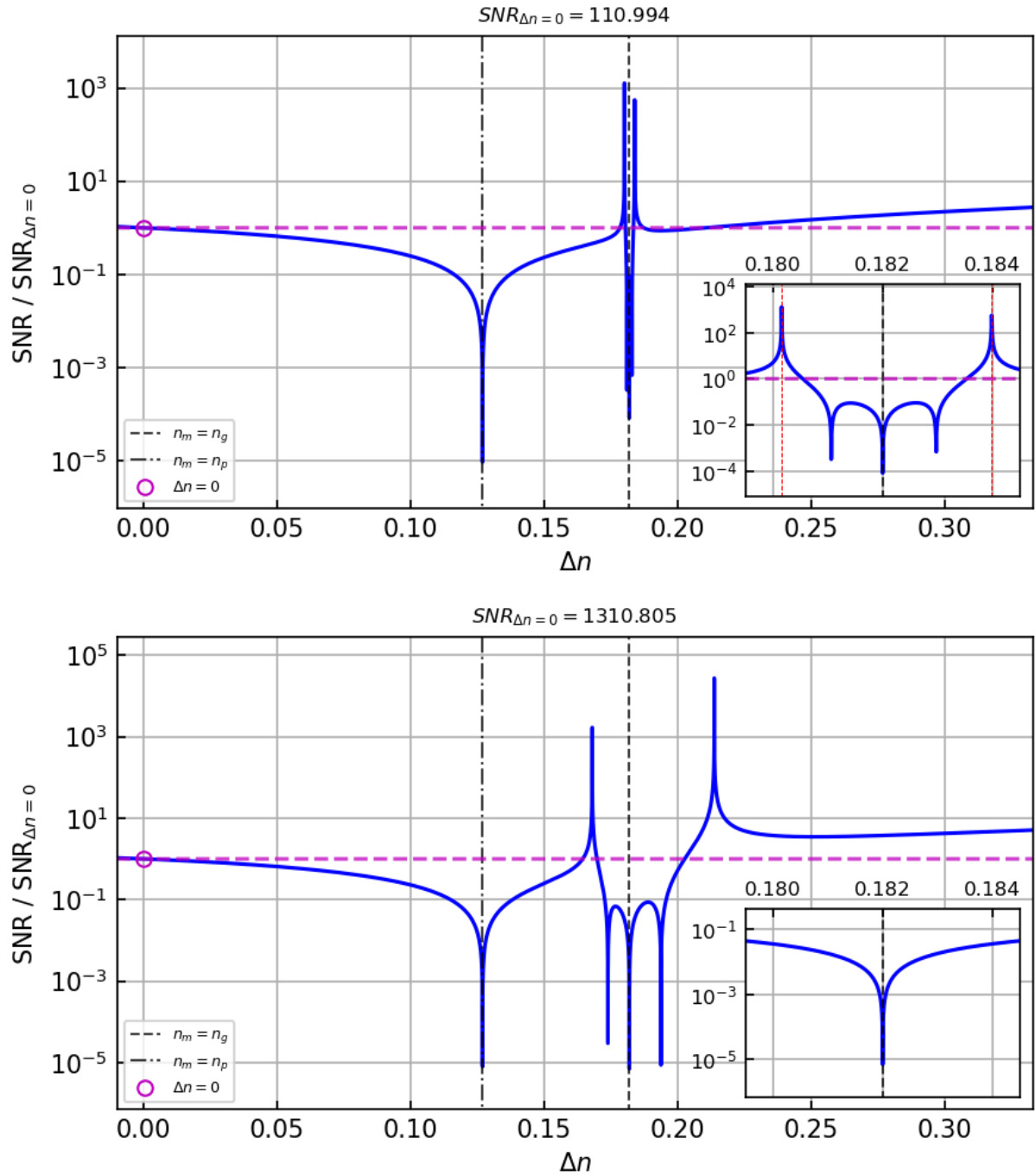

*Figure 3:* **Improvement of SNR through refractive index tuning.** *Top: Simulated SNR improvement w.r.t. $\Delta n = 0$ for BSA with a 0.1% mask. A strong SNR enhancement can be generated exactly where reference and scattering signals are equal (red dashed), albeit precise refractive index tuning is required. Bottom: Simulated SNR improvement w.r.t. $\Delta n = 0$ either for BSA with a 0.001% mask or 10x BSA with a 0.1% mask. Note that in these cases the same refractive index tuning as in Fig.1 & 2 would lead to a relative SNR reduction.*

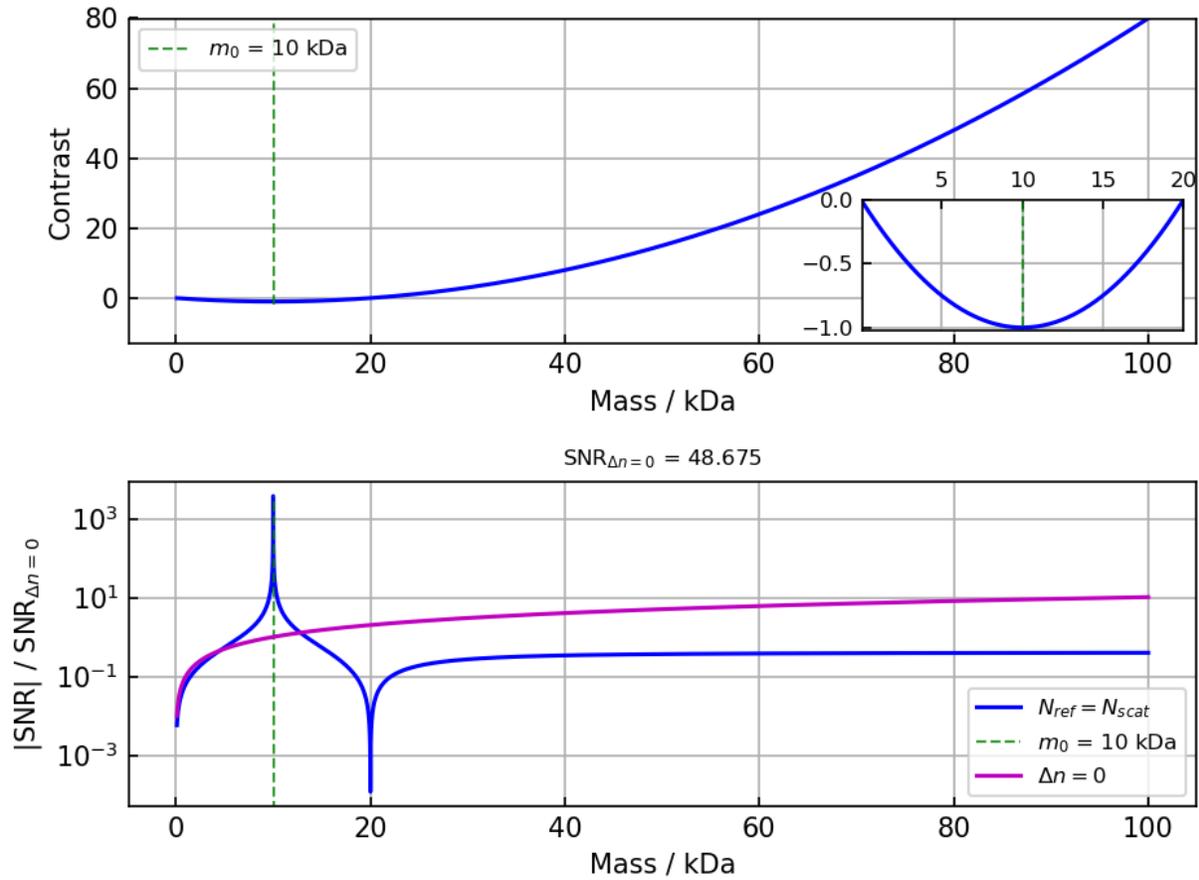

*Figure 4: **Inferring mass from the low-reference signal.** Top: Simulated quadratic contrast-to-mass conversion with minimum = -1 corresponding to the reference mass $m_0$ (here $m_0$ = 10 kDa). Bottom: Simulated SNR enhancement (blue) for different masses showing a strong improvement in the region close to $m_0$, w.r.t. $\Delta n = 0$ (magenta). Note the appearance of a blind spot at $m = 2 \cdot m_0$ and that particles below and above $m_0$, those that produce the same contrast, cannot be distinguished. The SNR enhancement < 1 (at $m > m_0$), is not detrimental as the absolute SNR of those larger mass species is still above the detection limit of standard mass photometry (SNR = 3; [5]).*